\documentclass[conference]{IEEEtran}
\IEEEoverridecommandlockouts
\usepackage{cite}
\usepackage{amsmath,amssymb,amsfonts}
\usepackage{algorithmic}
\usepackage{graphicx}
\usepackage{textcomp}
\usepackage{xcolor}
\usepackage{tabularx}
\usepackage{ragged2e}
\usepackage{caption}
\usepackage{textgreek}
\usepackage{float}
\usepackage{multirow}
\usepackage{subcaption}

\def\BibTeX{{\rm B\kern-.05em{\sc i\kern-.025em b}\kern-.08em
    T\kern-.1667em\lower.7ex\hbox{E}\kern-.125emX}}
\begin{document}
\title{IMS: Intelligent Hardware Monitoring System for Secure SoCs
\thanks{\textsuperscript{*} These authors contributed equally\\
This work has been partially funded  by the German Federal Ministry of Research, Technology and Space (BMFTR) through the project RILKOSAN.}
\thanks{This paper has been accepted for publication at the Design, Automation \& Test in Europe Conference (DATE) 2026.}
}

\author{\IEEEauthorblockN{Wadid Foudhaili\textsuperscript{1}\textsuperscript{*}, Aykut Rencber\textsuperscript{2}\textsuperscript{*}, Anouar Nechi\textsuperscript{1}, Rainer Buchty\textsuperscript{1}, Mladen Berekovic\textsuperscript{1},\\ Andres Gomez\textsuperscript{2}\textsuperscript{*}, and Saleh Mulhem\textsuperscript{1}\textsuperscript{*}}
\IEEEauthorblockA{\textit{\textsuperscript{1}Institute of Computer Engineering, Universit\"at zu L\"ubeck, L\"ubeck, Germany} \\
\textit{\textsuperscript{2}Institute of Computer and Network Engineering,TU Braunschweig, Braunschweig, Germany}}}

\maketitle

\begin{abstract}
In the modern Systems-on-Chip (SoC), the Advanced eXtensible Interface (AXI) protocol exhibits security vulnerabilities, enabling partial or complete denial-of-service (DoS) through protocol-violation attacks.
The recent countermeasures lack a dedicated real-time protocol semantic analysis and evade protocol compliance checks. 
This paper tackles this AXI vulnerability issue and presents an intelligent hardware monitoring system (IMS) for real-time detection of AXI protocol violations.
IMS is a hardware module leveraging neural networks to achieve high detection accuracy.
For model training, we perform DoS attacks through header-field manipulation and systematic malicious operations, while recording AXI transactions to build a training dataset.
We then deploy a quantization-optimized neural network, achieving 98.7\% detection accuracy with \textless=3\% latency overhead, and throughput of \textgreater2.5 million inferences/s.
We subsequently integrate this IMS into a RISC-V SoC as a memory-mapped IP core to monitor its AXI bus.
For demonstration and initial assessment for later ASIC integration, we implemented this IMS on an AMD Zynq UltraScale+ MPSoC ZCU104 board, showing an overall small hardware footprint (9.04\% look-up-tables (LUTs), 0.23\% DSP slices, and 0.70\% flip-flops) and negligible impact on the overall design's achievable frequency.
This demonstrates the feasibility of lightweight, security monitoring for resource-constrained edge environments.

\end{abstract}

\begin{IEEEkeywords}
AXI protocol security, hardware security monitoring, ML-based monitoring, SoC security, protocol-level attacks, denial-of-service, RISC-V.
\end{IEEEkeywords}

\section{Introduction}
Contemporary electronic devices have become ubiquitous, from smartphones to automotive systems. These devices are mainly powered by system-on-chip (SoC) architectures. Within these SoCs, individual components -- so-called Intellectual Property (IP) cores -- are interconnected and communicate using on-chip buses. Modern SoC architectures rely on the Advanced eXtensible Interface (AXI) protocol for high-performance communication between IP cores~\cite{SoCBusProtocols}. However, AXI's design prioritizes performance and flexibility over security, which inadvertently leads to security vulnerabilities and flaws.

Recent security analysis has revealed implementation flaws in AXI interconnects that enable new attack vectors targeting protocol-level vulnerabilities~\cite{Xray,Expect2025}. Analysis tools such as XRAY have identified 41 distinct implementation vulnerabilities in certain AXI interconnects~\cite{Xray,xilinx2022AXIviolations}, demonstrating that even protocol-compliant traffic can be exploited to bypass conventional protection mechanisms~\cite{Xray}. These vulnerabilities come from implementation flaws rather than protocol specification issues, yet enable sophisticated attacks~\cite{Expect2025,Xray}.

\subsection{SoC Denial of Service}
\begin{figure}[t]
    \centering
    \includegraphics[width=1\linewidth]{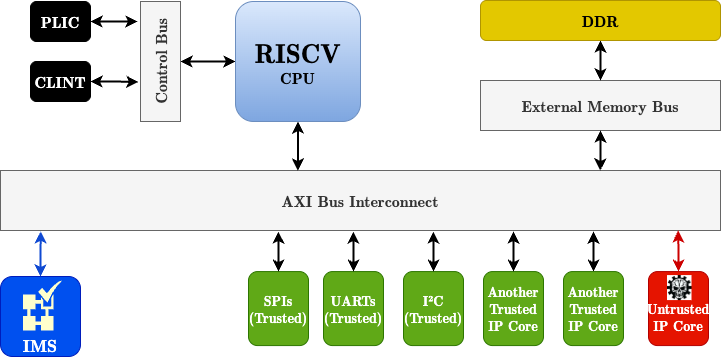}
    \caption{AXI-Bus Hardware Monitoring System Concept}
    \label{fig:IADS_Concept}
\end{figure}

Denial-of-service (DoS) attacks represent a critical attack vector where malicious or compromised components exploit implementation weaknesses and protocol characteristics to disrupt communication among SoC components~\cite{cocskun2024security}. Here, \textit{partial blocking} increases the unavailability of some IPs with a possibility of SoC malfunction, while \textit{complete blocking} causes immediate SoC malfunction. Fig.~\ref{fig:IADS_Concept} illustrates a typical SoC architecture where the host CPU connects to peripherals and IP cores via the AXI bus. Both malicious and buggy legitimate cores can trigger protocol violations, resulting in a partial or complete DoS. 
This can be performed or happens through malformed header fields, such as illegal burst lengths, transaction ID reuse, or signal flooding. Such protocol-level malicious operations remain undetectable to external monitoring systems because they occur within the SoC's internal communication fabric. Consequently, the SoC may continue operating while performance degrades, potentially violating service-level agreements or causing complete DoS.

The current security countermeasures and mechanisms focus primarily on access control, preventing unauthorized memory access through memory protection units~\cite{heinz2023dd,ewert2025lightweight} and interconnect policies~\cite{cocskun2024security,EarlySoCVP}. However, these approaches cannot detect protocol violations where malicious components use legitimate access patterns to violate protocol semantics through header field manipulation. This highlights a new class of SoC security challenges and vulnerabilities:
\begin{itemize}
    \item[C1] New attack vectors that exploit protocol semantics of on-chip buses.
    \item[C2] The inability of existing protection mechanisms to detect or mitigate such low-level real-time threats.
\end{itemize}

\subsection{Paper Contributions}
This paper presents a novel approach to detect and mitigate DoS attacks on SoC AXI buses using Machine Learning (ML).
We advance the state-of-the-art with the following key contributions:
\begin{itemize}
    \item We introduce a new security countermeasure called the intelligent hardware monitoring system (IMS) against critical AXI protocol-violation attacks, causing partial or complete DoS of the SoC. 
    \item We show how to build IMS as a machine learning model.
    Therefore, we start with the generation of a learning dataset, and we propose a thoroughly optimized hardware-friendly ML model by applying several ML optimization techniques.
    \item Our proposed IMS achieves a highly accurate detection rate of 98.7\%, making it suitable for real-world deployment.
    To demonstrate, we implement IMS in an RISC-V SoC and show the required hardware overhead for IMS integration.    
\end{itemize}
Our dataset for attack detection in AXI bus headers is made available for the public on our repository to allow reproducibility of the presented results and to address the lack of public datasets and benchmarks for protocol-level security research in the SoC environment. 

\section{Background \& Motivation}
\label{sec:related_work}
This section introduces our threat model and highlights its related state-of-the-art countermeasures. 
Then, we motivate our proposed solution.  

\subsection{Threat Model}
SoC security validation addresses the following security principles: Confidentiality, Integrity, Availability, and Authenticity~\cite{cocskun2024security}. 
This results in three main threat categories\cite{cocskun2024security}: (i) \textbf{Availability Violations:} Malicious or malfunctioning components/IP cores may make shared resources unavailable to legitimate users, (ii) \textbf{Confidentiality Breaches:} It mainly covers unauthorized access to sensitive data, and (iii) \textbf{Integrity Compromises:} Untrusted components try to modify a critical system state.

In this work, we focus on availability-violation threats, which represent a critical and immediate concern in SoC designs as they can cause immediate system-wide failures and are readily observable through performance degradation~\cite{Xray,xilinx2022AXIviolations}. While confidentiality and integrity violations are equally important in comprehensive security frameworks, availability attacks often serve as first indicators of system compromise and directly impact operational functionality.

Therefore, the proposed threat model considers an adversary who can be a malicious IP core or can exploit a malfunctioning IP to disrupt the availability of shared resources on the AXI bus infrastructure. 
This threat model is an extension of the established model in~\cite{EarlySoCVP}, which focuses on the transaction level of the AXI bus protocol~\cite{cocskun2024security,EarlySoCVP}.

\subsection{Hardware-Based Security Monitoring}

Current hardware security monitoring approaches for SoCs employ two primary strategies: \textbf{Memory Protection Unit (MPU)} and \textbf{Access Control Policy (ACP)}. 
MPU-based monitoring approaches enforce access boundaries to specific memory regions, ensuring only authorized components can read or write to them~\cite{heinz2023dd},while ACP-based monitoring approaches define communication rules among components/IP blocks using primarily address-based filtering~\cite{cocskun2024security,EarlySoCVP}.
However, these solutions focus on preventing unauthorized access rather than detecting protocol-level semantic violations.
Therefore, they exhibit fundamental limitations, as they rely on static threat models, cannot adapt to evolving attack patterns, and lack the intelligence to distinguish between legitimate transactions and protocol-compliant malicious behavior~\cite{TrustworthinessIC2024,TrustworthinessICSafety2024}.

\subsection{Research Gap and Motivation}
\label{sec:positioning}
Existing solutions focus on access control and memory protection but lack dedicated real-time protocol semantic analysis~\cite{Expect2025}. While static verification tools can identify design-time vulnerabilities, they cannot address runtime attacks that evade protocol compliance checks.
This indicates a critical security gap, which current countermeasures cannot overcome.
Therefore, we introduce an intelligent hardware monitoring system (IMS), deploying an ML model to monitor the AXI. 
ML has been intensively explored and investigated to monitor network transaction~\cite{Foudhaili2024Reconfigurable}, device operation, and CPU execution~\cite{EdgeAIBook2024}.    
Our work addresses this gap by introducing protocol-aware ML models deployed as a lightweight hardware engine for continuous on-chip monitoring.
To our knowledge, no existing approach employs ML for real-time AXI protocol monitoring in resource-constrained environments.

\section{IMS Design Methodology and Realization}
\label{sec:methodology}
This section presents the methodology for developing our intelligent IMS for real-time AXI protocol monitoring and security analysis. Our approach includes dataset generation, preprocessing pipeline design, ML model optimization and evaluation metrics.%
\begin{figure*}[t]
    \centering
    \includegraphics[width=1.77\columnwidth]{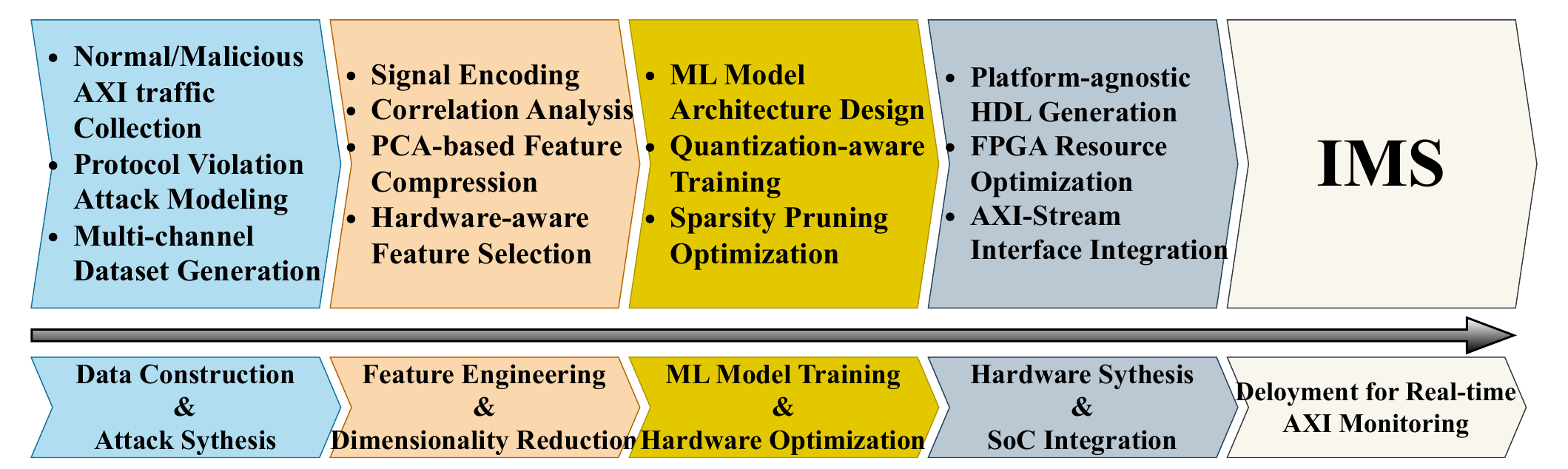}
    \caption{Overview of the IMS Design Steps}
    \label{fig:MethodologyADS}
\end{figure*}

\subsection{IMS Design Steps}
The key idea of our proposed work is designing an Intelligent Hardware Monitoring System (IMS) for AXI bus traffic.
The IMS operates as a passive hardware monitor strategically positioned within the AXI-bus interconnect to observe transaction patterns between system components without disrupting normal operation. 
The system architecture centers around a SoC interfacing with system resources through a hierarchical bus infrastructure and multiple devices (IP cores) competing for bus access via the AXI protocol, and hence worth use the AXI interconnect.
Untrustworthy components or IP cores, such as a malicious master or a malfunctioning device, can disrupt bus operation, potentially leading to system-wide stalls, and partial or complete DoS.
Therefore, our target is to design an IMS that can monitor the AXI-bus traffic in real time and analyze the AXI protocol. 
Fig.~\ref{fig:MethodologyADS} illustrates the IMS design steps. 
First, we build a training dataset by recording the transactions of the AXI bus traffic.
The dataset includes normal and malicious data. 
Indeed, malicious data indicates any abnormal behavior through header-level violations of the AXI protocol or resource starvation attacks. 
Then, we apply feature engineering methods on the dataset to use the most representative features during the training. 
We perform ML model optimization techniques and generate the RTL code for this ML model to serve as an IMS for any SoC. 

\subsection{Dataset construction and Attack Synthesis}

AXI has five channels: Write Address (AW), Write Data  (W), Write Response  (B), Read Address  (AR), and Read Data  (R), where %
anomalies can occur in any channel. 
The absence of publicly available AXI security datasets necessitated the creation of a comprehensive training corpus. 
We developed a dual-mode (~\textbf{Normal} and \textbf{Malicious operations}) data collection strategy using Chipyard Platform~\cite{Chipyard} for RISC-V SoC   
with standard AXI4 interconnect capabilities. 
This SoC includes multiple controllers (processors, DMA controllers, hardware accelerators) that act as AXI masters (initiators), competing for access to shared peripherals (e.g., memory, I/O), coordinated by the AXI interconnect’s arbitration logic.

\textbf{Normal Operation Capture:} We collected 16,383 legitimate transactions during standard Linux OS runtime, ensuring comprehensive coverage of typical system behavior across all five AXI channels (AW, W, R, B, AR). This baseline captures the natural transaction patterns expected in production SoC environments.
The normal transactions are captured during the System runtime, where nothing disturbs the transactions in any AXI channels.

\begin{figure}[h]
    \centering
    \begin{subfigure}{1\columnwidth}
        \includegraphics[width=1.04\columnwidth]{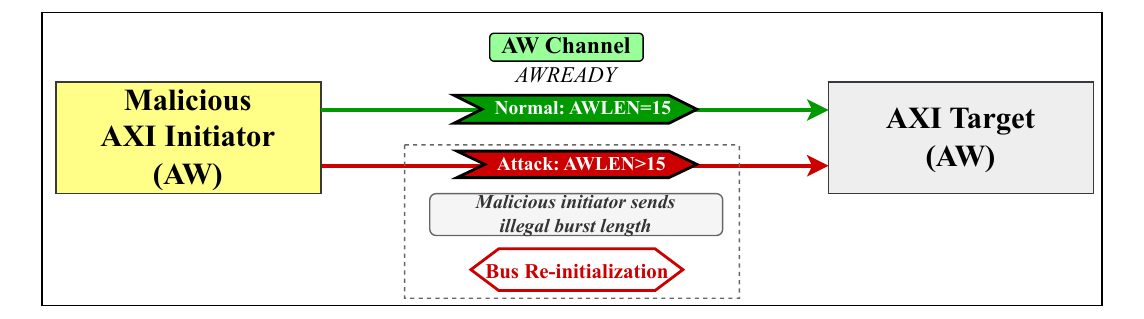}
        \caption{Attack 1: Illegal Burst Configuration}
        \label{fig:axi_attack1}
    \end{subfigure}

    \begin{subfigure}{1\columnwidth}
        \includegraphics[width=1.04\columnwidth]{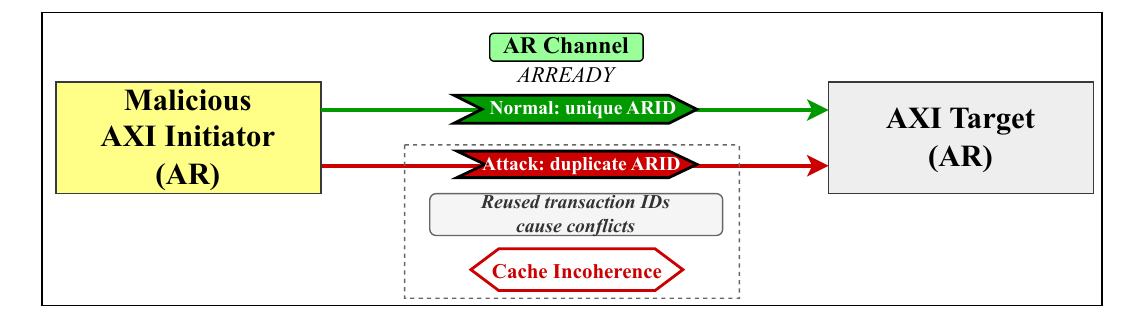}
        \caption{Attack 2: Transaction ID Exploit}
        \label{fig:axi_attack2}
    \end{subfigure}

    \begin{subfigure}{1\columnwidth}
        \includegraphics[width=1.04\columnwidth]{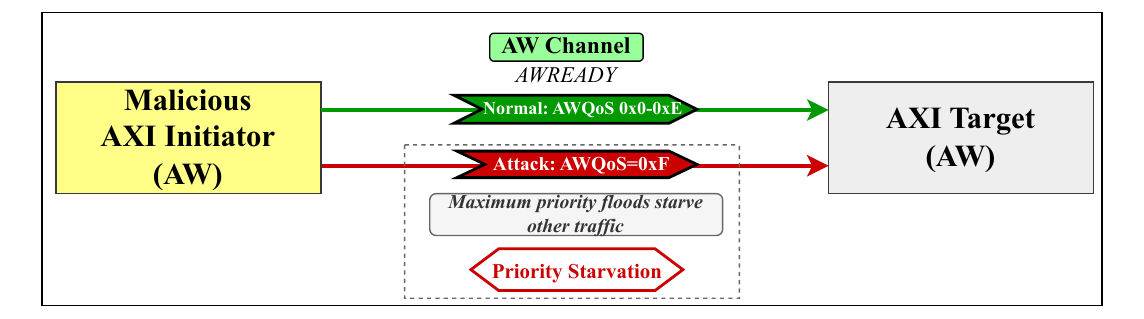}
        \caption{Attack 3: QoS Signal Flooding}
        \label{fig:axi_attack3}
    \end{subfigure}

    \includegraphics[width=1.06\columnwidth]{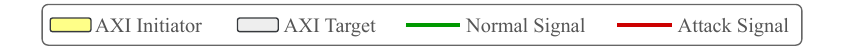}
    \caption{AXI Protocol Example of Security Vulnerabilities and Attacks Vectors
    }
    \label{fig:axi_attack_overview}
\end{figure}

\textbf{Malicious Operation Synthesis:} To record malicious operations, we introduce malicious behavior to the AXI protocol, covering several attack scenarios where a malicious or malfunctioning master (initiator) IP/component blocks the bus by not completing a write transaction, causing system-wide stalls and hardware-level DoS.
We systematically generated 3,242 attack samples representing three critical attack scenarios:
\begin{enumerate}
\item \textit{Illegal Burst Configurations [Fig.~\ref{fig:axi_attack1}]}: AWLEN values exceeding 15 to force bus re-initialization
\item \textit{Transaction ID Exploits [Fig.~\ref{fig:axi_attack2}]}: Duplicate ARID values inducing cache incoherence
\item \textit{QoS Signal Flooding [Fig.~\ref{fig:axi_attack3}]}: AWQOS saturation (0xF) to starve low-priority traffic
\end{enumerate}

These attacks lead to DoS of the targeted SoC. 

For instance, a malicious transaction can be injected as follows: After the initiator receives AWVALID in the AW channel, we hold WDATA in the W channel.      
These attack vectors simulate hardware-level scenarios where malicious initiators block bus transactions, causing system-wide stalls and a DoS.

Our instrumentation employs Vivado's Integrated Logic Analyzer (ILA) \cite{vivado_ila_UG908} with strategically placed probes at RISC-V initiator and AXI target/device interfaces to mirror all relevant AXI signals.%
We captured %
19,625 transactions (raw samples) divided into 16,383 normal and 3,242 malicious by performing the proposed three attack scenarios.
This provides sufficient data for robust ML training.
We export the captured data in three formats: raw, VCD, and CSV files, where the raw format is used for hardware instrumentation, VCD for dataset generation, and CSV format for ML-model training.
The raw dump file contains 290 signals corresponding to 57 features, represented in their simplest form as bits.

\subsection{Pre-processing Pipeline and Feature Engineering}

Our feature engineering methodology transforms raw protocol signals into machine learning-ready representations through a systematic three-stage pipeline designed to maximize information content while minimizing computational overhead.

(1)~\textbf{Stage 1 - Signal Encoding:} We decode binary and hexadecimal header fields to decimal representation using standardized AXI protocol specifications. After removing 5 ILA-specific debug signals not part of the AXI4 standard, 52 protocol-relevant features remain for subsequent processing. 
(2)~\textbf{Stage 2 - Correlation Analysis:} We apply statistical correlation analysis to identify and eliminate redundant features exhibiting high inter-correlation or constant values across samples. This process reduces the feature space from 52 to 22 dimensions, achieving a 58\% reduction while preserving essential discriminative information.

(3)~\textbf{Stage 3 - Dimensionality Optimization:} Principal Component Analysis (PCA) compresses the remaining features into lower-dimensional representations while retaining statistical significance. %
Our iterative approach achieves substantial compression: a combination of the original features as 4 components retains 90\% variance, 6 components preserve 95\%, and 8 components maintain 97\% of original information content.

The iterative application of correlation analysis and PCA continues until optimal feature reduction is achieved while maintaining the minimum acceptable variance threshold (90-97\%), ensuring that essential attack signatures remain detectable in the compressed feature space.

\begin{table}[h]
\caption{Dataset Statistics and Preprocessing Results}
\label{tab:datasetStatsPreproc}
\centering
\resizebox{\columnwidth}{!}{
\begin{tabular}{|l|c|c|c|}
\hline
\textbf{Metric} & \textbf{Normal} & \textbf{Malicious} & \textbf{Mixed} \\
\textbf{} & \textbf{} & \textbf{} & \textbf{Data} \\
\hline
Original Features & 52 & 52 & 52 \\
\hline
Post-Correlation Analysis (Features) & 22 & 22 & 22 \\
\hline
Post-PCA (90\% variance) (Components)  & 4 & 4 & 4 \\
\hline
Post-PCA (95\% variance) (Components) & 6 & 6 & 6 \\
\hline
Post-PCA (97\% variance) (Components) & 8 & 8 & 8 \\
\hline
\end{tabular}
}
\end{table}

\subsection{ML Model Architecture and Training}
We select a supervised Multilayer Perceptron (MLP) architecture based on its effectiveness with tabular protocol data and compatibility with hardware synthesis frameworks. Our network architecture comprises two hidden layers containing 32 neurons each with ReLU activation functions, followed by a sigmoid output layer optimized for binary classification. The model parameters are in Float32 data format.

\textbf{Training Configuration:} Data partitioning follows standard machine learning practices with 80\% allocated for training (augmented via SMOTE for class balance) and 20\% reserved for testing. We employ the ADAM optimizer with binary cross-entropy loss and L2 regularization (λ=0.0001) to prevent overfitting while maintaining generalization capability.

\textbf{Model Optimization:} Hardware deployment requirements necessitated aggressive optimization techniques. We implement QKeras-based~\cite{Qkeras} quantization-aware training, reducing weight precision to \textless8,5\textgreater format while applying 80\% sparsity pruning during training. These optimizations achieve substantial resource reduction without sacrificing detection accuracy, as validated through comprehensive performance metrics including accuracy, precision, recall, F1-score, and \textit{Area Under the Receiver Operating Characteristic Curve} (AUC-ROC) analysis.

\section{Hardware Implementation, Optimization and Discussion}
\label{sec:results}
This section presents the results of our experiments, including the performance of the model at different quantization levels, the results of the hardware implementation, and the performance of the attack detection. 
We also discuss the significance of these findings in the context of AXI protocol security.

\subsection{IMS Implementation and Optimization}
To implement the developed MLP as IMS on the targeted Chipyard SoC Platform~\cite{Chipyard}, we convert the trained MLP model into a hardware description language (HDL) using High-Level Synthesis (HLS) tools.
We use the HLS4ML\cite{fastml_hls4ml} framework to convert our trained model into HLS, resulting in a synthesized FPGA bitstream using the AXI-Stream interface.
The resulting IP core integrates seamlessly with existing SoC architectures through standardized AXI-Stream interfaces, enabling deployment as a memory-mapped peripheral or dedicated security coprocessor.

To enable efficient hardware mapping, further optimization is needed. 
We perform a QKeras-based\cite{Qkeras} weight quantization down to \textless8,5\textgreater precision to minimize memory footprint.
During training, we apply a constant sparsity pruning (target 80\%), forcing a fraction of zero weights to further reduce hardware usage. 
Table~\ref{tab:quantization_impact} shows the impact of quantization levels on ML performance. 
The \textless8,5\textgreater quantization level achieves optimal performance, even slightly outperforming the baseline Float32 model due to regularization effects. Performance remains acceptable until extreme quantization (\textless2,0\textgreater), where accuracy drops significantly. 
For this application, this shows the practical limits of weight compression. 
We chose the \textless8,5\textgreater  quantization (QKeras\cite{Qkeras}) in the experiments, as its recall score of 100\% shows a high detection rate.
We test its AUC-ROC, the result shows that the \textless8,5\textgreater quantization (QKeras\cite{Qkeras}) has AUC-ROC values exceeding 99\%. 

\begin{table}[!ht]
\caption{Quantization Impact on Model Performance}
\label{tab:quantization_impact}
\centering
\resizebox{.9\columnwidth}{!}{
\begin{tabular}{|l|c|c|c|c|}
\hline
\centering \textbf{Quantization} & \multirow{2}{*}{\textbf{Acc(\%)}} & \multirow{2}{*}{\textbf{Pr(\%)}} & \multirow{2}{*}{\textbf{R(\%)}} & \multirow{2}{*}{\textbf{F1(\%)}} \\
\centering \textbf{Level}       &                                   &                                &        &       \\
\hline
Float32 & 98.9 & 94.9 & 99.99 & 97.4 \\
\hline
\textless8.5\textgreater & 99.1 & 95.8 & 99.99 & 97.9 \\
\hline
\textless8.3\textgreater & 98.7 & 94.2 & 99.99 & 97.0 \\
\hline
\textless8.1\textgreater & 97.3 & 91.8 & 99.99 & 95.7 \\
\hline
\textless2.0\textgreater & 85.2 & 78.9 & 92.1 & 85.0 \\
\hline
\end{tabular}
}
\smallskip\\
\footnotesize{\textit{Note: Acc (Accuracy), Pr (Precision), R (Recall), F1 (F1 Score)}}
\end{table}

In our mapping of the quantized and pruned MLP to FPGA hardware (ZCU104), we aim to cover edge-device requirements with constrained resources.
Therefore, we focus on the following metrics: 
\begin{itemize}
    \item \textbf{Latency}, measured in milliseconds, quantifies the inference time per sample.
    \item \textbf{Throughput} is expressed by the number of inferences per second (inference/s).
    \item \textbf{Resource Utilization} results from the synthesis and elaboration of the hardware design, reporting the number of DSP slices, LUTs, flip-flops, and block RAM used.
\end{itemize}

Table~\ref{tab:res_hw_res} shows the hardware implementations on FPGA ZCU104. 
The quantized model achieves a noticeable reduction in resources, particularly in DSP slices (99.5\% reduction) and overall very low hardware resource usage on a ZCU104, making it suitable for resource-constrained SoC deployments. The low utilization percentages allow significant headroom for additional security features or multiple detector instances.

\begin{table}[!ht]
\caption{Detailed FPGA Resource Utilization (ZCU104)}
\label{tab:res_hw_res}
\centering
\begin{tabularx}{\columnwidth}{|m{1.3cm}|m{0.9cm}|m{1.2cm}|m{1.1cm}|m{1cm}|m{0.7cm}|}
\hline
\textbf{Resource} & \textbf{Baseline} & \textbf{Quantized} & \textbf{Reduction} & \textbf{HW} & \textbf{Usage} \\
\textbf{Type} & \textbf{Model} & \textbf{Model} & \textbf{(\%)} & \textbf{Resource} & \textbf{(\%)} \\
\hline
DSP Slices & 799 & 4 & 99.5 & 1,728 & 0.23 \\
\hline
LUTs & 45,238 & 20,841 & 53.9 & 230,400 & 9.04 \\
\hline
Flip-Flops & 8,450 & 3,224 & 61.8 & 460,800 & 0.70 \\
\hline
Block RAM & 12 & 8 & 33.3 & 312 & 2.56 \\
\hline
Clock Frequency & 250 MHz & 250 MHz & 0 & - & - \\
\hline
\end{tabularx}
\end{table}

To evaluate the performance of IMS, we increase AXI bus loads (10\%, 25\%, etc) while we measure the latency and the throughput of IMS. Table.~\ref{tab:PerformIMS} summarizes the IMS performance results, which show a stable IMS performance for different AXI bus loads.

\begin{table}[h]
\caption{IMS Performance Metrics vs. System Load}
\label{tab:PerformIMS}
\centering
\resizebox{.85\columnwidth}{!}{
\begin{tabular}{|c|c|c|}
\hline
\textbf{System Load} & \textbf{Inference Latency} & \textbf{Throughput} \\
\textbf{(\%)} & \textbf{(ms)} & \textbf{(inferences/s)} \\
\hline
10 & 1.523 & 2,567,891  \\
\hline
25 & 1.544 & 2,542,103 \\
\hline
50 & 1.566 & 2,509,578 \\
\hline
75 & 1.589 & 2,478,920 \\
\hline
100 & 1.612 & 2,445,682\\
\hline
\end{tabular}
}
\vspace{-0.5cm}
\end{table}

\subsection{IMS Detectability Analysis}

To perform such an analysis, we start by investigating and evaluating the overall IMS detection rate, and then we closely look at each attack.

\subsubsection{IMS Evaluation}
Table.~\ref{tab:model_evaluation} compares the performance of the baseline and quantized models. 
The high recall (R) score indicates that the IMS correctly detects the malicious operations. 
The high AUC-ROC score means that IMS can reliably separate malicious from normal operations.  
These results show an outstanding performance of IMS and its high level of detectability. 

\begin{table}[h]
\caption{IMS Performance Evaluation}
\label{tab:model_evaluation}
\centering
\resizebox{\columnwidth}{!}{
\begin{tabular}{|l|c|c|c|c|c|}
\hline
\textbf{Model} & \textbf{Acc} & \textbf{P} & \textbf{R} & \textbf{F1} & \textbf{AUC-ROC}\\
\textbf{} & \textbf{(\%)} & \textbf{(\%)} & \textbf{(\%)} & \textbf{(\%)} & \textbf{}\\
\hline
IMS (Baseline)& 98.9 & 94.9 & 99.99 & 97.4 & 0.993 \\
\hline
IMS (Quantized \textless8,5\textgreater) & 99.1 & 95.8 & 99.99 & 97.9 & 0.993 \\

\hline
\end{tabular}
}
\smallskip\\
\footnotesize{\textit{Note: Acc (Accuracy), Pr (Precision), R (Recall), F1 (F1 Score)}}
\end{table}

\subsubsection{Attack-Specific Detection Evaluation}
Table.~\ref{tab:attack_detection_performance} shows that the proposed IMS demonstrates consistently high detection rates across all attack types, with AWLEN-overflow attacks achieving perfect detection.
The performed attack-specific detection evaluation results in the following observations: 
\begin{itemize}
    \item \textbf{Burst-Length Exploits:} 100\% detection rate for illegal AWLEN values (\textgreater15) causing bus-stall attacks
    \item \textbf{QoS Flooding:} 97.3\% precision in detecting priority-inversion pattern attacks via abnormal AWQOS signals.
    \item \textbf{Transaction ID Reuse:} 98.1\% recall for cache-incoherence scenarios (ARIDS duplication identification)
\end{itemize}

Mixed attack patterns, representing real-world sophisticated attacks, maintain detection rates above 98\%.
The detection rate across the attacks validates the model's robustness against complex threat scenarios.

\begin{table}[h]
\caption{Attack-Specific Detection Performance Evaluation}
\label{tab:attack_detection_performance}
\centering
\begin{tabularx}{\columnwidth}{|m{3cm}|m{0.9cm}|m{0.5cm}|m{0.4cm}|m{0.8cm}|m{0.6cm}|}
\hline
\textbf{Attack Vector} & \textbf{Samples} & \textbf{T.P.} & \textbf{F.N.} & \textbf{D.R.(\%)} & \textbf{Pr(\%)} \\
\hline
AWLEN Overflow ($>15$) & 642 & 642 & 0 & 100.0 & 98.7 \\
\hline
ARID Duplication & 558 & 547 & 11 & 98.0 & 96.2 \\
\hline
AWQOS Flooding (0xF) & 423 & 411 & 12 & 97.2 & 94.8 \\
\hline
AWSIZE Invalid & 389 & 381 & 8 & 97.9 & 95.1 \\
\hline
ARRPROT Violation & 345 & 339 & 6 & 98.3 & 96.7 \\
\hline
Mixed Attack Pattern & 885 & 873 & 12 & 98.6 & 97.1 \\
\hline
Overall & 3,242 & 3,193 & 49 & 98.5 & 96.4 \\
\hline
\end{tabularx}
\smallskip\\
\footnotesize{\textit{Note: T.P. (True Positives), F.N. (False Negatives), D.R. (Detection Rate), Pr (Precision)}}
\end{table}

\begin{table*}[ht]
  \caption{Comparison of AXI Security \& Monitoring Solutions }%
  \label{tab:axi_comparison_enhanced}
  \setlength{\tabcolsep}{2.5pt} %
  \newcolumntype{Y}{>{\RaggedRight\arraybackslash}X}
  \begin{tabularx}{\textwidth}{@{}l l m{.7cm} m{1cm} m{1cm} m{1cm} m{1cm} m{1.4cm} l c c c@{}}
    \hline
    \textbf{Reference} & 
    \textbf{Target Prot.} & 
    \textbf{HW/ SW} & 
    \textbf{Timing Metrics} & 
    \textbf{Transac. Level} & 
    \textbf{Phase Level} & 
    \textbf{Protcol Check} & 
    \textbf{Header Process.$^\star$} & 
    \textbf{Perf. Metrics} & 
    \textbf{M.O. Supp.$^\dagger$} & 
    \textbf{Scalablility} \\
    \hline
    XRAY~\cite{Xray}                & AXI  & SW & -- & \checkmark & \checkmark & \checkmark & Static & -- & \checkmark & \checkmark \\
    \hline
    eXpect~\cite{Expect2025}        & AXI  & SW & -- & \checkmark & \checkmark & \checkmark & Static & -- & \checkmark & \checkmark \\
    \hline
    TMU Tiny~\cite{liang2025towards} & AXI4 & HW & \checkmark & \checkmark & -- & -- & Basic & (error logs) & \checkmark & \checkmark \\
    \hline
    TMU Full~\cite{liang2025towards} & AXI4 & HW & \checkmark & \checkmark & \checkmark & -- & Basic & \checkmark & \checkmark & \checkmark \\
    \hline
    DD-MPU~\cite{heinz2023dd}       & generic & HW & -- & \checkmark & -- & -- & Addr-only & -- & -- & \checkmark \\
    \hline
    VAST IFT~\cite{cocskun2023vast} & VP Level   & SW & -- & -- & -- & -- & N/A & -- & -- & \checkmark \\
    \hline
    Trust MU~\cite{trustworthSoC}   & AHB/ Wishbone  & HW & \checkmark & \checkmark & -- & -- & Basic & behavior & -- & -- \\
    \hline
    \textbf{This work}        & AXI4 & HW & \checkmark & \checkmark & \checkmark & \checkmark & \textbf{ML-based} & DSP, Latency, Throughput & \checkmark & \checkmark (ML) \\
    \hline
  \end{tabularx}
  \vspace{1ex}\\
  {\footnotesize 
  $^\star$\textbf{Header Process.} indicates the depth of protocol header field analysis: \textit{Static} = design-time verification only; \textit{Basic} = simple rule-based validation; \textit{Addr-only} = address-based filtering; \textit{ML-based} = intelligent runtime header analysis that detects malformed headers appearing as valid transactions.\\
  $^\dagger$M.O.\ Supp.\ denotes Multiple Outstanding transaction (initiated by not completed) support.}
\end{table*}

Our findings indicate that the IMS exhibits perfect detection (100\% rate) against very critical protocol violation attacks (AWLEN overflow) while maintaining high detection rates (\textgreater97\% precision) across all attack vectors, including sophisticated mixed attack patterns and a false positive rate far below 1\% (at 99\% recall).
This demonstrates our model's robustness and reliability in real-world scenarios.

\section{Related Work}
Recently, AXI monitoring has become a crucial task, especially for anomaly and malicious operation detection purposes.
Such monitoring increases the SoC availability against AXI DoS and arouses interest in the context of security. 

To show the advantages of our proposed IMS, we systematically compare it with the state-of-the-art solutions, shown in Table.~\ref{tab:axi_comparison_enhanced}.
Although XRAY\cite{Xray} and eXpect\cite{Expect2025} serve as robust static analysis frameworks, offering significant protocol coverage at both transaction and phase levels during the design phase. 
Nonetheless, these software-centric strategies present notable operational limitations, particularly their inability to address runtime threats or adapt to the dynamic attack patterns. 
This static nature prevents the detection of sophisticated runtime exploits that manifest through protocol-compliant malicious operations.
The Transaction Monitoring Unit (TMU) variants proposed~\cite{liang2025towards} exhibit hardware-based real-time monitoring capabilities with TMU Full\cite{liang2025towards} providing comprehensive coverage across transaction and phase-level monitoring. 
However, these monitoring solutions fundamentally rely on simple rule-based validation methods that lack a semantic comprehension of protocol violations. 
TMU Tiny\cite{liang2025towards} achieves minimal overhead through simplified error logging, yet it fails to recognize intricate attack patterns. 
Conversely, while TMU Full\cite{liang2025towards} offers broader coverage, it remains fundamentally limited in its ability to differentiate between legitimate transactions and semantically malicious behaviors that comply with protocol standards. 
The DD-Memory Protection Units (DD-MPU)\cite{heinz2023dd} illustrates memory-centric security strategies that focus solely on address-based filtering mechanisms. 
DD-MPU achieves efficiency and scalability in hardware implementation, however, it operates independently of the protocol-level semantic violations, which are critical attack vectors. The address-only processing paradigm can not detect changes (manipulations) in header fields and shows a limited security countermeasure against illegal burst configurations or QoS flooding attacks.
Vast IFT \cite{cocskun2023vast} and Trust MU \cite{trustworthSoC} are specialized frameworks that address distinct protocol contexts. 
These solutions demonstrate efficiency within their targeted domains but cannot address the challenges of AXI protocol monitoring.
Our proposed IMS uniquely processes AXI protocol-header features using an ML-based approach. Unlike static verification frameworks which operate exclusively at design time, or basic rule-based hardware monitors, our IMS intelligently detects sophisticated attack patterns appearing as legitimate transactions to conventional validation systems.

\section{Conclusion}
\label{sec:conclusionFW}

The protocol-level violations against AXI bus used in the system-on-chip (SoC)
leads to a system-on-chip's (SoC) partial or complete denial of service. 
This work presents the first comprehensive approach for detecting protocol-level malicious operations by using a machine learning (ML) model integrated into RISC-V-based SoCs as a memory-mapped IP core. 
We address the absence of public datasets for low-level SoC-bus violations by constructing a novel dataset built from captured normal AXI traffic and from systematically injected DoS, both collected via a RISC-V SoC on a Xilinx ZCU104 FPGA platform.
The optimized ML model achieves high accuracy (up to 99.11\% AUC-ROC) and robust performance in precision, recall, and F1 metrics. The model is deployed as an IP core on the FPGA and demonstrates real-time inference with minimal resource usage (0.23\% DSP and 0.70\% Flip-flops utilization on a ZCU104), low latency (1.566 ms), and high throughput (\textgreater2.5 million inferences/s).
This work demonstrates that protocol-aware, ML-based detection of malicious operations can be effectively realized in resource-constrained edge environments, providing a practical path for securing SoC interconnects against sophisticated hardware attacks.\par 

As future work, we aim to investigate the following avenues:
\begin{itemize}
    \item Broader attack coverage: Extend the attack scenarios to cover additional AXI channels and include protocol violations beyond DoS scenarios
    \item ML Model diversity: Investigate alternative ML architectures (e.g., LSTM, GNN) for zero-day attack detection.
    \item Integration with SoC security frameworks: Explore co-design with existing SoC firewalls, hypervisors, or runtime monitors to provide layered, defense-in-depth protection.
    \item ASIC integration using state-of-the-art technology libraries.
\end{itemize}

\newpage
\bibliographystyle{ieeetr}
\bibliography{Bibliography/conf}
\end{document}